\documentstyle[preprint,eqsecnum,aps]{revtex}

\begin{document}
\title{Noncommutative Geometric Gauge Theory from Superconnections \\}
\author{Chang-Yeong {\sc Lee}\\} 
\address{Department of Physics, Sejong University, Seoul
    143-747\\
E-mail: leecy@phy.sejong.ac.kr\\}
\maketitle

\hspace*{.5cm}

\begin{abstract}
Noncommutative geometric gauge theory is 
reconstructed based on the superconnection concept. The bosonic action of 
the Connes-Lott model including the symmetry breaking Higgs sector 
is obtained
by using a new generalized derivative, which consists of
the usual 1-form exterior derivative plus an extra element called 
{\it the matrix derivative}, for the curvatures.
We first derive the matrix derivative based on superconnections and then 
show how the matrix derivative can give rise to spontaneous 
symmetry breaking.
We comment on the correspondence
between the generalized derivative and the generalized Dirac operator
of the Connes-Lott model.
\end{abstract}


\pagebreak

\section{Introduction\protect\\}

Several years ago, Connes \cite{conn} proposed to use 
noncommutative geometry for particle physics models. 
Then he and Lott \cite{conlot}
showed that one could obtain the standard model from a noncommutative 
geometric gauge theory. 
In this noncommutative framework, the Dirac K-cycle 
plays an important role. For physicists,
however, this formalism demands quite a bit of mathematics. 
Here, we present a formalism, which we hope is relatively easier 
for physicists
to understand, based on the superconnection concept \cite{qui,ns}. 
Coquereaux and other people \cite{couqet,sch,pps,ht} prosposed models for 
a noncommutative geometric gauge theory
in which they also used a matrix commutator operator similar to the one that
we use in this paper. Especially, the term {\it matrix derivative} was coined 
in one of these works \cite{sch}.
However, these works were based on different concepts,
as we shall mention later.

In this paper, we first review the Connes-Lott approach to
noncommutative geometric gauge theory for self-containedness.
Then we present the superconnection approach: After a brief introduction
to superconnections, we derive the matrix derivative by 
generalizing the exterior
derivative, write the Yang-Mills action with curvatures construructed
from the generalized derivative including the matrix derivative,
 and show how 
{\it the matrix derivative} can give rise to symmetry breaking.
Finally, we comment on the correspondence between our 
generalized derivative and the generalized Dirac operator of the Connes-Lott
approach
and compare our approach with others and then make conclusions.

\section{Connes-Lott Approach\protect\\}

In the Connes-Lott approach, the Dirac K-cycle on a * (involution) 
algebra acting
on a Hilbert space plays an important role. Both spacetime and internal
space are described by the involution algebra, and 
the newly introduced generalized Dirac operator is crucial  
for symmetry breaking, which is related to 
the fermionic mass matrix. 

A K-cycle on a * algebra ${\cal A}$ is given by \\ 
$\bullet$  a faithful representation $\rho$ of ${\cal A}$ by bounded 
operators on a Hilbert space ${\cal H}$, \\ 
$\bullet$  a self-adjoint (generalized Dirac) operator 
${\cal D}$ on ${\cal H}$
 such that $[{\cal D}, a]$ is a bounded operator for all $a \in {\cal A}$,
and $(1 + {\cal D}^2 )^{-1}$ is a compact operator on ${\cal H}$. \\
In the Connes-Lott approach, 
an even K-cycle is used, which has, in addition, \\ 
$\bullet$  a self-adjoint ${\bf Z}_{2}$ grading (chirality) 
operator $\Gamma$ on ${\cal H}$ such that $\Gamma^2 =1, \; 
\Gamma {\cal D} + {\cal D} \Gamma = 0$, and $\Gamma a = a \Gamma$ for
all $ a \in {\cal A}$.
This even K-cycle is called {\it the Dirac K-cycle}.

Another important ingredient of the Connes-Lott approach is the
$\pi$ representation of the universal differential 
envelop of ${\cal A}$, $\Omega^{*}({\cal A})$, where
$\Omega^{*}({\cal A}) = \oplus \Omega^{k}({\cal A})$ such that
$\Omega^{0}({\cal A})={\cal A}$ and 
$\Omega^{k}({\cal A})= \{ a_{0}\delta a_{1} \cdots \delta a_{k}; \;
 a_{0}, a_{1}, \cdots, a_{k} \in {\cal A} \}$, the space of universal
k-forms. 
The differential $\delta$ satisfies 
$ \delta^2 =0, \; \; \delta(a_{0}\delta a_{1} \cdots \delta a_{k})=
\delta a_{0}\delta a_{1} \cdots \delta a_{k} \in \Omega^{k+1}({\cal A}),$
and the involution * is given by
$(a_{0}\delta a_{1} \cdots \delta a_{k})^* =\delta a_{k}^* \cdots 
\delta a_{1}^* a_{0}^* .$
Now, $\pi$ is a map from  $\Omega^{*}({\cal A})$ to ${\cal B (H)}$, the
space of bounded operators on ${\cal H}$, given by
\begin{equation}
 \pi(a_{0}\delta a_{1} \cdots \delta a_{k})=\rho(a_{0})
[{\cal D}, \rho(a_{1})] \cdots [{\cal D}, \rho(a_{k})] . 
\end{equation}
Note that, in order to respect the nilpotency of $\delta$, ${\cal D}$ 
should satisfy $ [{\cal D}, [{\cal D}, \ \cdot \ ] ] =0 .$

Finally, the tensor product of two noncommutative spaces with Dirac
K-cycles, $({\cal A}_{1},{\cal H}_{1},{\cal D}_{1},\Gamma_{1})$ and 
$({\cal A}_{2},{\cal H}_{2},{\cal D}_{2},\Gamma_{2})$, is defined as
\begin{eqnarray}
 {\cal A}&=&{\cal A}_{1}\otimes {\cal A}_{2}, \; \;
 \; \; \;   {\cal H}={\cal H}_{1}\otimes {\cal H}_{2}, \nonumber \\ 
{\cal D}&=&{\cal D}_{1} \otimes 1 + \Gamma_{1}\otimes {\cal D}_{2} \; \;
     \; ({}^{\rm or} = {\cal D}_{1} \otimes\Gamma_{2} +
    1 \otimes {\cal D}_{2} ), \label{c2} \\
   \Gamma &=&\Gamma_{1}\otimes \Gamma_{2}, \; \; \; \; \; 
  \Omega^{*}({\cal A})=\Omega^{*}({\cal A}_{1}) \otimes 
    \Omega^{*}({\cal A}_{2}). \nonumber 
\end{eqnarray}
Here, the definition of ${\cal D}$ in the product space depends on
which of the  initial spaces corresponds to spacetime. 
For instance, if space 1 corresponds
to spacetime, then we use the first definition; if space 2
does, then we use the second one.
Now, we get into the models of the Connes-Lott approach.

\subsection{Two-point space:}

Take ${\cal A}= {\bf C} \oplus {\bf C},$ 
${\cal H}= {\bf C}^N \oplus {\bf C}^N,$ and
${\cal D}= \left( \begin{array}{cc} 0 & M^{\dag} \\
 M  & 0  \end{array} \right),$  $\Gamma=\left( \begin{array}{cc} 
 1 & 0 \\ 0  & -1  \end{array} \right)$ where $M$ is an $N\times N$ 
matrix. If $a=(\lambda, \lambda') \in {\cal A}$ then
$\pi(\delta a)= [{\cal D}, \rho(a)]= (\lambda -\lambda')
\left( \begin{array}{cc} 0 & - M^{\dag} \\
 M  & 0  \end{array} \right)$. 
The connection is given by
${\cal J} = a_{0} \delta a_{1}$ with $ a_{0}=(u,u'), \; 
a_{1}=(v,v') \in {\cal A}$; thus, $\pi({\cal J})=\pi(a_{0} \delta a_{1})=
\rho(a_{0})[{\cal D}, \rho(a_{1})]=(v-v') \left( \begin{array}{cc} 
0 & -u M^{\dag} \\ u' M  & 0  \end{array} \right).$ From  
$ {\cal J} = a_{0} \delta a_{1}=(u,u')\cdot (v'-v,v-v')
=(u(v'-v),u'(v-v'))$ and denoting it by ${\cal J}\equiv
(\phi^* , \phi )$, we can write $\pi({\cal J})=\left( \begin{array}{cc}
0 & \phi^*  M^{\dag} \\ \phi  M  & 0  \end{array} \right)$.
The $\pi$ representation of the curvature $\theta$ is given by
$\pi(\theta)=\pi(\delta {\cal J} + {\cal J}^2)=(\vert \phi +1 \vert^2
-1) \left( \begin{array}{cc}
M^{\dag}M  & 0 \\ 0  & MM^{\dag}   \end{array} \right).$
Now the Yang-Mills action is given by
$I_{\bigtriangledown}={\rm Tr}_{\omega}((\pi(\theta))^2 
{\cal D}_{\bigtriangledown}^{-n})$ where ${\rm Tr}_{\omega}$ is
the Dixmier trace, $n$ is the dimension of the manifold, and 
${\cal D}_{\bigtriangledown}$ is the ``covariant derivative" given by 
${\cal D}_{\bigtriangledown}={\cal D} + \pi({\cal J})$.
Since $n$ is zero and the Dixmier trace becomes the
usual trace in the present case, the Yang-Mills action is given by
$I_{\bigtriangledown}=2 (\vert \phi +1 \vert^2 -1)^2 
{\rm Tr} (M^{\dag}M)^2 .$ This is just the Higgs potential with minima
at $\phi =0, -2$; this type of potential indicates 
explicitly broken symmetry. 

\subsection{Spinmanifold: }

Take  ${\cal A}= C^{\infty}(Z) \otimes {\bf C}$ where $Z$ is a 4-dimensional
spinmanifold, and let ${\cal H}= L^2(S)$ where $S$ is the vector bundle of
spinors on $Z$. Then, the Dirac and the chirality operators become the usual 
ones, ${\cal D}= \gamma^{\mu}\partial_{\mu}$ and $\Gamma=\gamma_{5}$.
The connection is an ordinary differential 1-form on $Z$, $ \pi({\cal J})=A$.
Thus, the curvature is given as the usual $\pi(\theta)=F=dA +A^2$, 
and the Yang-Mills action is 
$I_{\bigtriangledown}=\int_{Z} {\rm Tr}(F*F)$. 

\subsection{Product space:}
 	 
Consider now the tensor product of the above two spaces. Following the
given tensor product rule of Eq. (\ref{c2}), we get
\begin{eqnarray}
{\cal A}& = & C^{\infty}(Z)\otimes ({\bf C}\oplus{\bf C}), \; \;
{\cal H}= L^2(S)\otimes ({\bf C}^N \oplus {\bf C}^N), \nonumber \\
{\cal D}&=& \gamma^{\mu}\partial_{\mu} \otimes I_{2} \otimes I_{N} +
   \gamma_{5} \otimes \left( \begin{array}{cc} 0 & M^{\dag} \\
 M  & 0  \end{array} \right), \\
  \Gamma &= &\gamma_{5}\otimes 
\left( \begin{array}{cc} 1 & 0 \\ 0  & -1  \end{array} \right)
\otimes I_{N}. \nonumber
\end{eqnarray}   
The connection is given by $ \pi({\cal J})=\left( \begin{array}{cc}
A & \phi^* \gamma_{5} M^{\dag} \\ \phi \gamma_{5} M  & A'
  \end{array} \right)$. The curvature is thus given by 
\begin{eqnarray*}
\pi(\theta) & = & \pi ( \delta {\cal J} + {\cal J}^2) \\
            & = & \left( \begin{array}{cc} dA +AA +(\phi + \phi^* +
       \phi^* \phi)M^{\dag}M & (d \phi^* + A\phi^* -\phi^* A' +A -A')
      \gamma_{5} M^{\dag}\\ 
 (d \phi - \phi A + A' \phi -A + A') \gamma_{5} M  & 
 dA' +A'A' + ( \phi + \phi^* + \phi \phi^*)M M^{\dag}  \end{array} \right)
\end{eqnarray*}
where $d= \gamma^{\mu} \partial_{\mu}$ 
 and $A=\gamma^{\mu} A_{\mu}, \; A'=\gamma^{\mu} {A'}_{\mu}$ are a 1-form 
exterior derivative and gauge fields, respectively.
The Yang-Mills action now mainly consists of three parts, the usual 
Yang-Mills action, the kinetic part of scalar field, and the 
Higgs potential: 
\begin{equation}
 I_{\bigtriangledown} \sim \int_{Z}{\rm Tr}( \alpha_{1} (\vert F_{A}\vert^2 +
 \vert F_{A'}\vert^2) + \alpha_{2}\vert D \phi \vert^2 +
\alpha_{3}(\vert \phi +1 \vert^2 -1)^2 + \cdots ) 
\end{equation}
where $D \phi = d \phi + A' \phi - \phi A $ and ``$\cdots$" denotes 
both mass terms for $A, \ A'$ and interaction terms. 

\section{Superconnection Approach\protect\\}

The superconnection was first introduced in mathematics by Quillen in
1985 \cite{qui}. However, in physics this concept was used earlier
in 1982 by
Thierry-Mieg and Ne'eman without giving it a name \cite{tmn} 
under the notion of a generalized connection {\it a la} Cartan \cite{crt}. 
Then in 1990, Ne'eman and Sternberg \cite{ns} used superconnections 
for a Higgs
mechanism in a manner for physicists to understand this concept much
easier than that of Quillen's.
We now follow the Ne'eman-Sternberg presentation of superconnections 
in this paper.
 
Let $V = V^{+} \oplus V^{-}$ be a super (or $Z_{2}$-graded) complex vector
space; then, the algebra of endomorphisms of $V$ is a superalgebra with
even or odd endomorphisms.
Let ${\cal E}={\cal E}^{+} \oplus {\cal E}^{-}$ be 
a super (or $Z_{2}$-graded) vector bundle over a manifold $M$, and
$\Omega(M) =\oplus \Omega^k(M)$ be the algebra of smooth differential
forms with complex coefficients.
Then, let $\Omega(M, {\cal E})$ be the space of ${\cal E}$-valued 
differential foms on $M$. 
This space has a $Z \times Z_{2}$ grading; however, we are
mainly concerned with its total $Z_{2}$ grading. Hence, $\Omega(M, {\cal E})$ 
can be regarded as a supermodule over $\Omega(M)$.
The total $Z_{2}$ grading can be denoted by 
$\Omega(M, {\cal E})=\Omega^+ (M, {\cal E}) + \Omega^- (M, {\cal E})$
which is defined by
\[ \Omega^{\pm} (M, {\cal E})=\sum_{k}\Omega^{2k} (M, {\cal E}^{\pm}) \oplus 
\sum_{k} \Omega^{2k + 1} (M, {\cal E}^{\mp}) \]
where $2k \; (2k+1)$ indicates an even (odd) exterior form degree. 
Now we consider the tensor product of
$\Omega(M)$ and
${\cal A}\equiv {\rm End}(V)$, which belongs to
${\rm End} (\Omega(M, {\cal E}))$.
We decompose ${\cal A}$ as ${\cal A}= {\cal A}^+ \oplus
{\cal A}^-$ such that
 ${\cal A}^+$ consists of all ``matrices" of the form
\[  \left( \begin{array}{cc} R & 0 \\
 0  & S  \end{array} \right), \; \; \; R \in {\rm End}(V^+), \;
  \; \; S \in {\rm End}(V^-), \] while  ${\cal A}^-$ consists of all 
``matrices" of the form
\[  \left( \begin{array}{cc} 0 & K \\
 L  & 0  \end{array} \right), \; \; \; K \in {\rm Hom}(V^-, \ 
 V^+), 
\; \; \; L \in {\rm Hom}(V^+, \ V^-). \] 
If we choose bases of $ V^+$ and $ V^-$, then we can think of 
$R, \ S, \ K,$ and $L$ as actual matrices. We can, thus, think of elements of
$\Omega(M)\otimes {\cal A}$ as matrices whose entries are 
differential forms.
For instance, both $ \left( \begin{array}{cc} \omega_{0} & 0 \\
 0  & \omega_{1}  \end{array} \right) $ and 
$ \left( \begin{array}{cc} 0 & L_{01} \\
 L_{10}  & 0  \end{array} \right) $
are odd elements of $\Omega(M)\otimes {\cal A} $ if
$\omega_{0}, \ \omega_{1}$ are matrices of odd degree differential forms
and  $ L_{01}, \  L_{10}$ are matrices of even degree differential forms.

A superconnection $\bigtriangledown$ on ${\cal E}$ is an odd element of 
${\rm End} (\Omega (M, {\cal E}))$, that is, 
\begin{equation}
 \bigtriangledown \ : \; \Omega^{\pm} (M, {\cal E}) \longrightarrow
   \Omega^{\mp} (M, {\cal E}) , 
\end{equation}
and satisfies the derivation property
\begin{equation}
 \bigtriangledown (v \alpha)=( dv)\alpha +
  (-1)^{\vert v \vert} v \bigtriangledown \alpha, \;
 \; \; v \in \Omega (M), \; \; \alpha \in \Omega (M, {\cal E}) 
\end{equation} 
where 
 $d$ is  a 1-form exterior derivative operator which is odd, 
and $\vert v \vert$ is the exterior degree of $v$. 
In terms of the local trivialization of ${\cal E}$, 
say ${\cal E}= M \times V$
(locally), the most
general superconnection can be written locally as 
\begin{equation}
 \bigtriangledown =  d + \omega , \; \; \; 
   \omega \in (\Omega (M) \otimes{\cal A})^- = \Omega^+ (M) \otimes
   {\cal A}^- \oplus \Omega^- (M)\otimes {\cal A}^+ . 
\end{equation} 
In ``matrix" language, $d$ is given by  ${\bf d} =\left( 
\begin{array}{cc} d & 0 \\ 0  & d  \end{array} \right) $ 
with $d$ inside
the matrix denoting the usual 1-form exterior derivative operator
times a unit matrix, and $\omega$ is given by
$\omega = \left( \begin{array}{cc} \omega_{0} & L_{01} \\
  L_{10}  & \omega_{1}  \end{array} \right) $ where 
$\omega_{0}, \ \omega_{1}$ are matrices of odd degree differential forms
and  $ L_{01}, \  L_{10}$ are matrices of even degree differential forms. 
Here, the multiplication rule is given by
\begin{equation}
 (u \otimes a)\cdot (v \otimes b) = (-1)^{\vert a \vert  \vert v \vert}
 (uv)\otimes (ab), \; \; \; u,v \in \Omega(M), \; \; a,b \in {\cal A}. 
\end{equation}  
In what follows, we will use this matrix language.

We now derive the way in which {\it the matrix derivative} 
enters into play in this
superconnection formulation.
First, we note that the 1-form exterior derivative ${\bf d}$ is nilpotent and
satisfies the derivation rule
\begin{equation}
 {\bf d}^2 =0, \; \; \;
{\bf d}(\alpha \beta) = ({\bf d} \alpha) \beta + (-1)^{\vert \alpha \vert}
      \alpha ({\bf d} \beta), \; \; \; \alpha, \beta \in 
  \Omega (M, {\cal E}) 
\end{equation}  
where $\vert \alpha \vert$ is the total $Z_{2}$ grading of $\alpha$. 
Next, we want to generalize this operator acting on $ \Omega (M, {\cal E})$,
say ${\bf d}_{t}$,
 keeping the above two properties such that 
\begin{equation}
 {\bf d}_{t}^2=0, \; \; \;
{\bf d}_{t}(\alpha \beta) = ({\bf d}_{t} \alpha) \beta + 
(-1)^{\vert \alpha \vert} \alpha ({\bf d}_{t} \beta), \; \; \; \
 \alpha, \beta \in \Omega (M, {\cal E}).
\end{equation} 
To make things easy, we look for an object that can be added to ${\bf d}$ as
a component of ${\bf d}_{t}$, and write this additional component as
${\bf d}_{M}$: $ \; {\bf d}_{t}= {\bf d}+{\bf d}_{M}$.
Thus, the two conditions that ${\bf d}_{t}$ should satisfy now become
\begin{eqnarray}
  \; \; {\bf d}_{M}^2 &=& 0, \; \; \; \; \; \; \; 
   {\bf d}{\bf d}_{M} + {\bf d}_{M}{\bf d}=0, \label{c11} \\
 {\bf d}_{M}(\alpha \beta)& =& ({\bf d}_{M} \alpha) \beta +
(-1)^{\vert \alpha \vert} \alpha ({\bf d}_{M} \beta), \; \; \; \
 \alpha, \beta \in \Omega (M, {\cal E}). \nonumber
\end{eqnarray}
Since ${\bf d}_{M}$ should behave as a part of the superconnection operator
\cite{bgv} in a sense, we write it as a (graded) commutator operator 
\begin{equation}
{\bf d}_{M}= \left[ \eta, \; \cdot \ \right], \; \; \; \eta \in 
   \Omega (M, {\cal E}) 
\end{equation} 
and fix $\eta$ such that ${\bf d}_{M}$ satisfies the conditions
in Eq. (\ref{c11}).
The derivation rule and the condition $ {\bf d}{\bf d}_{M} + 
{\bf d}_{M}{\bf d}=0$ require that $\eta$ be odd and ${\bf d} \eta =0$;
i.e. $ \vert \eta \vert = 1$ and the elements of $\eta$ should be
 closed forms.
The nilpotency condition $ {\bf d}_{M}^2=0$ is satisfied if $\eta^2$
commutes with any element in $\Omega (M, {\cal E})$. 
There are two simple choices satisfying this condition: \\
$\bullet \; \; \;  \eta =   \left( \begin{array}{cc} u & 0 \\
 0  & v  \end{array} \right) $ where $u, \; v$ are odd degree closed forms
with their coefficient matrices satisfying $ u^2 = v^2 \propto 1$, or \\
$\bullet \; \; \; \eta =   \left( \begin{array}{cc} 0 & m \\
 n  & 0  \end{array} \right)$ where $m, \; n$ are even degree closed forms
with their coefficient matrices satisfying $mn=nm \propto 1$. \\
In this paper, we choose the second one with 0-form $m, \; n$.
We express this choice by ${\bf d}_{M}= [\eta , \ \cdot \ ] $
with $ \eta =
 \left( \begin{array}{cc} 0 & \zeta \\ \overline{\zeta}  & 0  
\end{array} \right)$ where $  \zeta,  \; \overline{\zeta}$ are 
 0-form constant matrices satisfying $  \zeta \overline{\zeta} =
\overline{\zeta} \zeta \propto 1 $.
We call this operator ${\bf d}_{M}$  {\it the matrix derivative}
following the terminology used in Ref. 6.

With the use of this generalized derivative, the curvature is now given by
\begin{equation}
 {\cal F}_{t}=({\bf d}_{t} + \omega )^2 = {\bf d}_{t} \omega 
   + \omega^2 .  
\end{equation}
In this formulation, we write the Yang-Mills action as
\begin{equation}
 I_{t} = \int_{M} {\rm Tr} ( {\cal F}_{t}^{\star} {\cal F}_{t} ) 
\end{equation}
where $\star$ denotes taking dual for each entry of ${\cal F}_{t}$ 
in addition to taking the Hermitian conjugate.  
In the next section, it will  be shown that this action will provide
the Yang-Mills-Higgs action with spontaneously broken symmetry that 
we obtained in the previous section through the Connes-Lott approach.

\section{Symmetry Breaking and Comparison between the Two 
 Approaches\protect\\}

In order to compare the action $I_{t}$ obtained in the previous section
 with the one given in Section II, we assign
0-form scalar fields in the odd part and 1-form
gauge fields in the even part of $\omega$ of the superconnection:
\begin{equation}
 \omega = \left( \begin{array}{cc} A & \phi^* \\
 \phi  & A'  \end{array} \right) . 
\end{equation}
Now, we choose $ \eta =  \left( \begin{array}{cc} 0 & 1 \\
 1  & 0  \end{array} \right) $ for simplicity.
Then, the curvature is given by
\begin{equation}
 {\cal F}_{t} = \left( \begin{array}{cc} F_{A} + (\phi^*  \phi
 + \phi + \phi^*) & D \phi^* + A - A' \\
 D \phi + A' - A  & F_{A'} + ( \phi \phi^* +
    \phi + \phi^* )
  \end{array} \right) 
\end{equation}
 where 
$ F_{A}=dA +A^2, \; \ F_{A'}=dA' + {A'}^2, \; \
 D \phi = d \phi + A' \phi - \phi A ,$ and
$ D \phi^* = d \phi^* + A \phi^* - \phi^* A' .$
Thus, the Yang-Mills action in the superconnection approach is given by
\begin{eqnarray}
 I_{t}& =& \int_{M}{\rm Tr} ( {\cal F}_{t}^{\star} {\cal F}_{t} ) \\
        & \sim & \int_{M}{\rm Tr} \left[ \vert F_{A}\vert^2 +
\vert F_{A'}\vert^2 + 2(\vert D \phi \vert^2 +
 (\vert \phi +1 \vert^2 -1)^2 ) + \cdots \right] \nonumber 
\end{eqnarray}
where ``$\cdots$" includes the mass terms for $A, \; A'$ and 
interaction terms.
Now, the Higgs potential clearly shows that symmetry is broken explicitely
\cite{couqet,lhn,thl}. This action is the same as the one we obtained in 
the product space case of the Connes-Lott approach in Section II.

Now, we would like to compare the generalized derivative in the 
superconnection approach
with the generalized Dirac operator in the Connes-Lott approach.
The generalized Dirac operator in the product space case in Section II
was given by 
\begin{equation}
{\cal D}=\left( \begin{array}{cc} d & 0 \\
 0  & d  \end{array} \right) +
   \left( \begin{array}{cc} 0 &\gamma_{5} M^{\dag} \\
  \gamma_{5} M  & 0  \end{array} \right), 
\end{equation}
and the differential $\delta$ translates
into a commutator operator, $\ [ {\cal D}, \ \cdot \ ], $
in the $\pi$ representation.
In the superconnection approach, the 1-form exterior derivative can be
written as ${\bf d}= [ {\bf d}, \ \cdot \ ]$ since
\begin{equation} 
 [{\bf d}, \alpha ] = {\bf d}\alpha -\alpha {\bf d}=({\bf d} \alpha ), \; \;
   \alpha \in  \Omega (M, {\cal E}).
\end{equation}
Therefore, the generalized derivative ${\bf d}_{t}={\bf d} +{\bf d}_{M}$
can be expressed as 
\begin{equation}
{\bf d}_{t}= [{\bf d}, \ \cdot \ ] +{\bf d}_{M}.
\end{equation}
Remember that ${\bf d}=\left(
\begin{array}{cc} d & 0 \\ 0  & d  \end{array} \right) $
and ${\bf d}_{M}=[\eta , \ \cdot \ ]$ where 
$\eta =\left( \begin{array}{cc} 0 & \zeta \\ \overline{\zeta}  & 0
\end{array} \right)$.
Thus, we can see that the generalized Dirac operator in the two-point
space case plays the role of the matrix derivative in the superconnection
approach. Also, the name {\it fermionic mass matrix} for $M$ in the two-point
space case will be clear if we look into the fermionic action 
\begin{equation}
I_{\Psi} = \int_{Z} \overline{\Psi} {\cal D}_{\bigtriangledown} \Psi 
\end{equation}
where ${\cal D}_{\bigtriangledown}={\cal D} + \pi({\cal J})$.
If we write $\Psi=(\psi_{L},\psi_{R})$, then the generalized Dirac operator
induces the mass terms $ \overline{\psi}_{L}\gamma_{5} M^{\dag} \psi_{R}
+ \overline{\psi}_{R}\gamma_{5} M \psi_{L} $.
In the superconnection approach, the fermionic action is given by
\begin{equation}
I_{\Psi} =\int_{M} \overline{\Psi} ({\bf d}_{t} + \omega) \Psi, \; \; \;
\Psi \in V \otimes S   
\end{equation}
where $S$ is a spinor bundle.
Since $V$ is a $Z_{2}$-graded (super) vector space, we can write
$\Psi=(\psi_{+},\psi_{-})$. Then the matrix derivative term induces
the mass terms, $\overline{\psi}_{+} \zeta \psi_{-} +
\overline{\psi}_{-} \overline{\zeta} \psi_{+}$, 
in this case also.

\section{Comments and Conclusions\protect\\}

As we have seen, the role of the matrix derivative in the 
symmetry-breaking
mechanism is very similar to that of the generalized Dirac operator. 
Without introducing the generalized Dirac operator, 
whose matrix component is called
the fermionic mass matrix, it was not possible to include the scalar
field as a part of the connection in the Connes-Lott approach.
Also, as its component name, {\it fermionic mass matrix},  suggests, 
the generalized 
Dirac operator is essential for spontaneous symmetry breaking in that
 approach. Similarly,
although the scalar field is a part of the superconnection from the
beginning in the superconnection approach, there would be no 
``automatic" symmetry breaking without the matrix derivative \cite{ln,hln}.

Our approach looks similar to some approaches presented earlier 
\cite{couqet,sch} in the sense that there too the ``matrix derivative"
was used
in conjunction with Connes-Lott's two-point space idea. Especially, 
in one of them, Ref. 6, the term ``matrix derivative" was coined.
However, in those works, this matrix commutator operator
was simply defined to conform with 
the generalized Dirac operator and to induce the Connes-Lott action,
as their main architect Coquereaux confirms \cite{ccq}. 
Thus, the scalar field was regarded as a 1-form 
in Coquereaux {\it et al.}'s approach \cite{ccq}, as it should be 
in the Connes-Lott approach, 
 rather than as the 0-form that it should be in the superconnection approach.

We also note that the notion of a $Z_{2}$-graded (super) vector space 
which is
essential to the superconnection construction was crucial for the
correspondence between this approach and the Connes-Lott approach. 
This is because in the Connes-Lott approach the two-point space structure
is very basic in obtaining the standard model 
since it induces the left and the right movers even if one deals with 
a more complicated structure \cite{is}. 
In the superconnection approach this requirement
is provided by the $Z_{2}$-graded vector space \cite{krf}. 

In conclusion, here we have shown the way in which
 the generalization of the exterior
derivative in the superconnection framework can provide the so-called
matrix derivative, and we have produced the Yang-Mills-Higgs action 
with built-in
symmetry breakdown mechanism, which was obtained from the Connes-Lott model
through the generalized Dirac operator.
\\

\acknowledgments 

I would like to thank Dae Sung Hwang and Yuval Ne'eman for 
collaborations on related subjects. Thanks are also due to 
Hoil Kim and Youngwon Kim for detailed explanations on various 
mathematics concepts related to the subject. This work was supported 
in part by the 1994 grant from Daeyang Academic Foundation. 
\\

\end{document}